\documentclass{article}
\usepackage{graphicx} 
\usepackage{url}
\usepackage{hyperref}
\usepackage{cite}
\usepackage[utf8]{inputenc}
\usepackage[font=footnotesize,labelfont=bf]{caption}
\usepackage{color}
\usepackage{tabularx}
\usepackage{array}
\usepackage{float}
\usepackage{amsmath}
\usepackage{amssymb}
\usepackage{tablefootnote}

\usepackage[a4paper,top=3cm,bottom=3cm,left=2.5cm,right=2.5cm,bindingoffset=0mm]{geometry}

\newcolumntype{C}[1]{>{\centering\arraybackslash}p{#1}}

\begin{document}

\begin{center}


{\Large \bf QED Effects in PDFs \\ \vspace*{0.3cm} - A Les Houches Comparison Study}

\vspace*{1cm}
 Thomas Cridge$^{a,b}$, Juan Cruz Martinez$^{c}$, Joey Huston$^{d}$ \\                                 
\vspace*{0.5cm}    

\end{center}

\begin{minipage}{0.9\textwidth}
\begin{center}
$^a$ Elementary Particle Physics, University of Antwerp, Groenenborgerlaan 171, 2020 Antwerp, Belgium \\
$^b$ Department of Physics and Astronomy, University of Manchester, Manchester, M13 9PL, United Kingdom \\
$^c$ Departamento de Física Atómica, Molecular y Nuclear, Facultad de Física, Universidad de Sevilla, E-41080 Sevilla, Spain \\
$^d$ Department of Physics and Astronomy, Michigan State University, East Lansing, MI 48824, USA \\
\end{center}
\end{minipage}

\begin{abstract}
In the last decade, and even more so in the last few years, our knowledge of the internal structure of the proton has become more accurate and precise thanks to the large amount of data available and developments in theory and methodology.
The reduction of the uncertainties associated to these developments has brought previously neglected effects into focus as their typical magnitude are competitive with the size of the uncertainties.
One such effect is the inclusion of QED into PDF fits. Typically this is a percent effect, and thus while theoretically important, it has had a relatively limited impact on phenomenological studies up to this point.
In this proceeding we study some of the effects which, while peripheral to the inclusion of QED in the proton, can considerably change the relative size and shape of the QCD+QED fit with respect to the QCD only determination. These may become important in the future as precision continues to increase. After a comparison of the QCD+QED PDFs with the QCD only PDFs of various global PDF fitting groups, we focus largely upon NNPDF4.0, which shows the biggest effect when including QED. Focusing largely on a single set of PDFs also enables more subtle effects to be analysed, making it an ideal candidate for this study.
\end{abstract}

\section{Introduction}

Parton Distribution Function (PDF) determination in the past few years has reached ever-improving levels of accuracy and precision. The combination of higher order QCD theoretical inputs, with the community standard having now reached NNLO~\cite{Bailey:2020ooq,NNPDF:2021njg,Hou:2019efy,ATLAS:2021vod,Alekhin:2017kpj} and approximate N3LO (aN3LO)~\cite{McGowan:2022nag,NNPDF:2024nan}, with increasing amounts of precision LHC data has allowed PDF fitters to reach PDF uncertainties at the level of a few percent or better. As a result, a variety of small previously-neglected effects must be considered, from the inclusion of missing higher order theoretical uncertainties~\cite{McGowan:2022nag,NNPDF:2024dpb}, to higher twist corrections~\cite{Alekhin:2012ig,Cerutti:2025yji,Harland-Lang:2025wvm,Ball:2025xtj}, to the impact of QED and electroweak effects~\cite{Xie:2021equ,Xie:2023qbn,Cridge:2021pxm,NNPDF:2024djq,Cridge:2023ryv,Barontini:2024dyb}. In this work we focus on the latter, in particular upon the impact of including QED into the DGLAP evolution and the resulting generation of a photon PDF on the partonic distributions relative to the QCD only case. 

The development of the LUXQED formalism~\cite{Manohar:2016nzj,Manohar:2017eqh}, relating the photon PDF of the proton to precisely measured structure functions measured in deep inelastic scattering data, has enabled substantial improvement in the determination of the photon PDF. As a result, all major global PDF-fitting groups now provide QCD+QED PDFs\footnote{We use QCD+QED to describe the PDFs with QED effects included in general, however in the literature these are often referred to as simply QED PDFs, which we use occasionally when they are the names the groups use for such PDFs.} on top of their default QCD baselines at NNLO~\cite{Xie:2021equ,Xie:2023qbn,Cridge:2021pxm,NNPDF:2024djq}, and also at aN3LO in the case of MSHT and NNPDF~\cite{Cridge:2023ryv,Barontini:2024dyb}. Naturally, the inclusion of such QED effects represents a better approximation to the true underlying physics. Nonetheless, unlike the case in QCD where there have been substantial benchmarking efforts between the global PDF fitting groups in the past~\cite{Cridge:2021qjj,PDF4LHCWorkingGroup:2022cjn}, such a benchmarking has not yet been performed for QED impacts on the PDFs. 

The lack of such a benchmarking of QED effects in PDFs motivated work to compare the QED-induced PDF effects across the global fitting PDF groups in~\cite{Cooper-Sarkar:2025sqw}, to which our investigations here contributed and build upon. It was observed that whilst the different PDFs consistently determine the same total photon momentum (of $0.4\%$ at $Q=M_H = 125~\mathrm{GeV}$), there are subtle differences in the impact this has on the momentum distributions of the other (QCD) partons. The net effect of the addition of the photon PDF is to reduce the momentum in the other partons, as required by the momentum sum rule, with reductions in the total gluon and quark singlet momenta at $Q=M_H = 125~\mathrm{GeV}$ found to be in the range $0.4-0.8\%$ and $0.1-0.4\%$ respectively~\cite{Cooper-Sarkar:2025sqw}. While these changes impact the cross-sections for several processes, in \cite{Cooper-Sarkar:2025sqw} Higgs production in a variety of channels was investigated. In particular, the reductions in the gluon momenta were found to have important consequences for Higgs production in gluon-fusion. Overall, this resulted in the recommendation in~\cite{Cooper-Sarkar:2025sqw} to include this reduction via a straightforward average of the impacts seen by the three groups entering PDF4LHC21~\cite{PDF4LHCWorkingGroup:2022cjn}.

\section{Comparison of Gluon-Gluon Luminosities}

\begin{figure}[H]
\begin{center}
\includegraphics[scale=0.45,trim=0cm 0cm 0cm 1.395cm,clip]{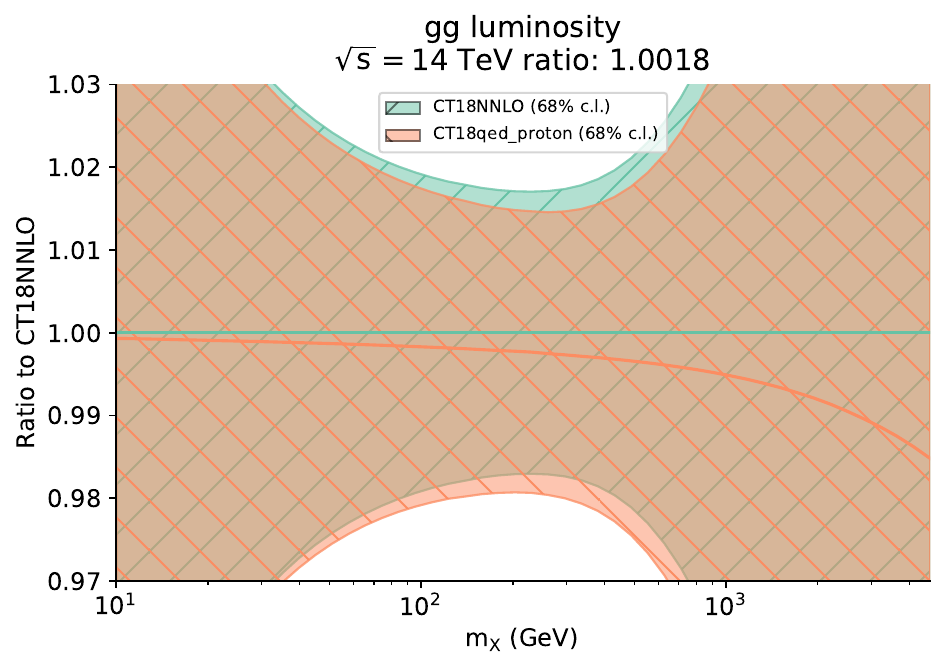}
\includegraphics[scale=0.45,trim=0cm 0cm 0cm 1.395cm,clip]{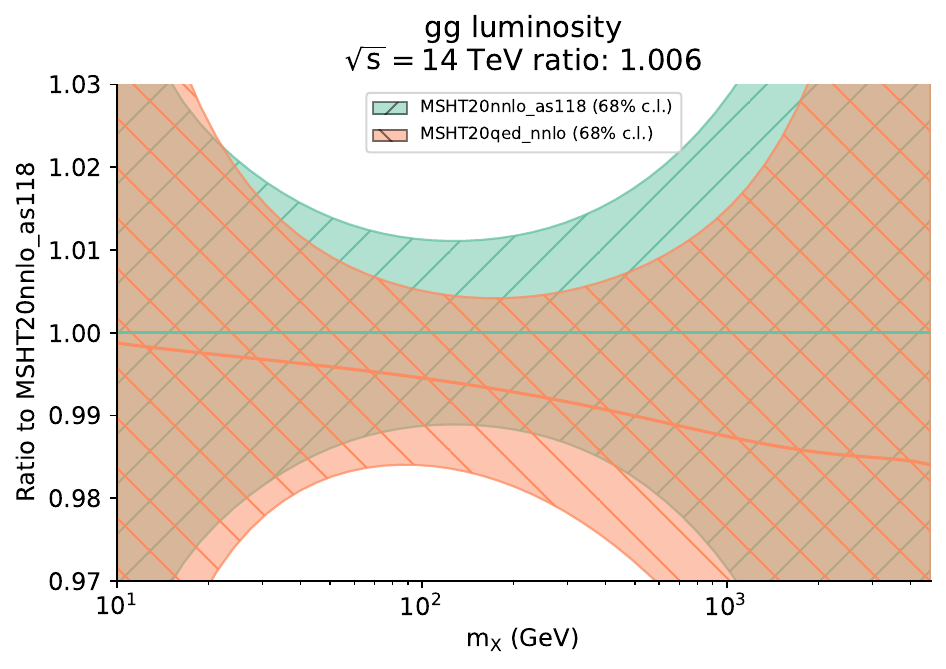}
\includegraphics[scale=0.45,trim=0cm 0cm 0cm 1.395cm,clip]{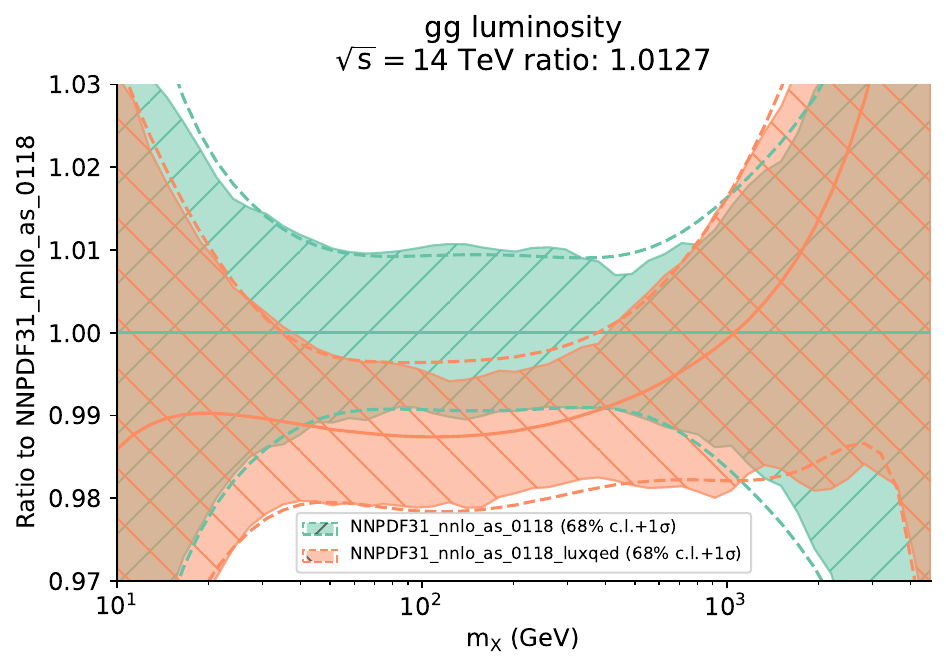}
\includegraphics[scale=0.45,trim=0cm 0cm 0cm 1.395cm,clip]{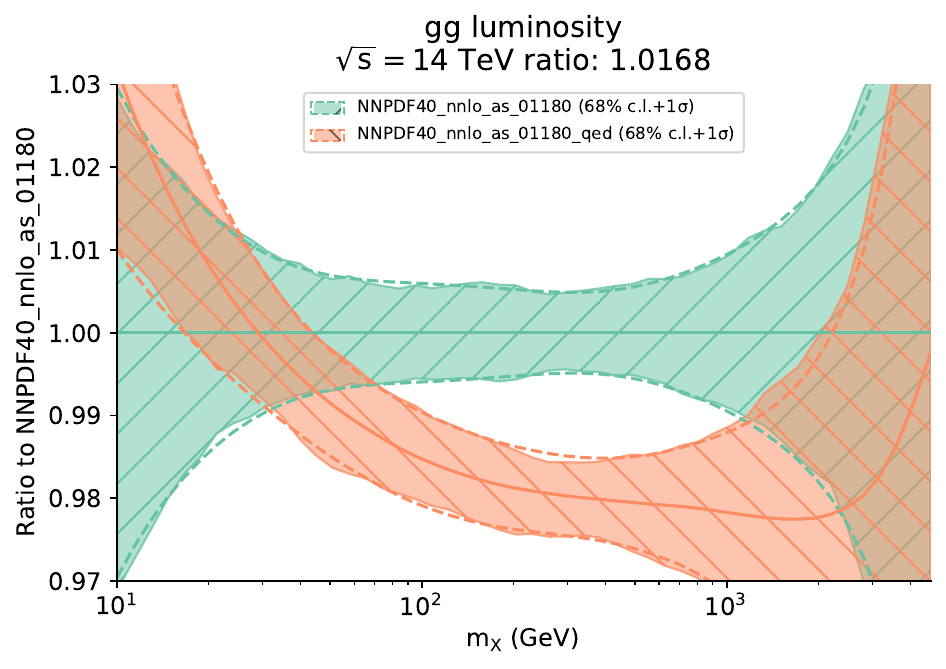}
\caption{\sf Comparison of the gluon-gluon luminosities at $\sqrt{s}=14~\rm{TeV}$ for the QCD+QED and QCD only default PDFs of CT18, MSHT20, NNPDF3.1 and NNPDF4.0 at NNLO.}
\label{fig:gglumis_defaultNNLOQEDPDFs}
\end{center}
\end{figure}

As part of these studies the gluon-gluon luminosity at $m_H = 125~\rm{GeV}$ for $\sqrt{s}=14 ~\rm{TeV}$ was computed for a variety of different QCD only and QCD+QED PDF sets~\cite{Hou:2019efy,Xie:2021equ,Xie:2023qbn,Bailey:2020ooq, Cridge:2021pxm,McGowan:2022nag, Cridge:2023ryv,NNPDF:2017mvq,NNPDF:2021njg,NNPDF:2024djq,NNPDF:2024nan,Bertone:2017bme,NNPDF:2024djq,Barontini:2024dyb,Ball:2025xgq}. We begin by comparing the versions of the PDFs of the different groups available on LHAPDF in  Table~\ref{tab:1}. Whilst all QCD+QED PDFs observe a reduction in the gluon-gluon luminosity in Table~\ref{tab:1}, a range of results across the different PDF sets is observed. The gluon-gluon luminosities as a function of the invariant mass, $m_X$ are also presented in Figure~\ref{fig:gglumis_defaultNNLOQEDPDFs}, where different shapes as well as magnitudes of effects are observed. A question to address is therefore the reason for such differences and the extent to which the differences are related to differences in the implementation of the QED effects in the PDFs or are rather intrinsic to the fits themselves, i.e. related to differences also present in the QCD baselines such as from methodology or data included. For the former of these, whilst all three groups implement the LUXQED formalism, several subtle differences and choices that can be made were pointed out in \cite{Cooper-Sarkar:2025sqw}. These range from the manner in which the photon momentum is obtained from the pure QCD partons (e.g. whether this is determined by the fit, or enforced by-hand from certain specific partons) to the scale at which the photon PDF is extracted. For the latter, differences in the baseline QCD PDFs - in terms of data included, methodology and in the approach to the inclusion of electroweak corrections can also influence the effects observed from the addition of QED effects.

\begin{table}[h!]
\fontsize{8}{10}\selectfont 
  \renewcommand\arraystretch{1.1} 
\centering
\begin{tabular}{l C{1.2cm} C{1.4cm} C{2.4cm} C{2.65cm}}
\hline
\textbf{PDF set} & \textbf{Central} &
\textbf{\% error (symmetrised)} &
\textbf{\% change rel. to QCD only baseline at same order} &
\textbf{\% change rel. to NNLO QCD baseline} (3.1 for NNPDF) \\
\hline
MSHT20NNLO & 0.0206 & 1.16 & -- & -- \\
MSHT20NNLO+QED & 0.0205 & 1.09 & -0.60 & -0.60 \\
MSHT20aN3LO & 0.0195 & 1.71 & 0 & -5.20 \\
MSHT20aN3LO+QED & 0.0193 & 1.47 & -0.99 & -6.14 \\
\hline
CT18 & 0.0205 & 1.93 & -- & -- \\
CT18+QED proton & 0.0205 & 1.93 & -0.19 & -0.19 \\
\hline
NNPDF3.1 & 0.0210 & 1.06 & -- & -- \\
NNPDF3.1+QED & 0.0207 & 0.74 & -1.36 & -1.36 \\
NNPDF4.0 & 0.0207 & 0.56 & 0 & -1.44 \\
NNPDF4.0+QED & 0.0204 & 0.52 & -1.64 & -3.05 \\
NNPDF4.0aN3LO (inc. MHOU) & 0.0202 & 0.65 & 0 & -3.85 \\
NNPDF4.0aN3LO+QED (inc. MHOU) & 0.0198 & 0.55 & -1.88 & -5.63 \\
\hline
\end{tabular}
\caption{PDF $gg$ luminosities for $\sqrt{s}=14$~TeV at $m_H=125$~GeV. The default versions of the PDFs available on LHAPDF are shown.}
\label{tab:1}
\end{table}

The purpose of this proceedings is to elucidate these differences and we investigate a selection of these in this work. We will begin in Section~\ref{sec:NNLO} investigating this at NNLO, before moving on to aN3LO in Section~\ref{sec:N3LO}.

\subsection{NNLO}\label{sec:NNLO}

In order to look further into these differences in the LHAPDF sets' QED impacts amongst the groups we expand Table~\ref{tab:1} to include the greater range of PDF sets provided in the groups' QED studies, we focus on NNLO in Table~\ref{tab:2}. 

For CT18 we include also the additional alternative treatments of the QCD+QED PDFs provided by the CT18 LUX and CT18 QED fit PDF sets~\cite{Xie:2021equ,Xie:2023qbn} in addition to the standard CT18 QED proton set. These three PDF sets with QED corrections explore the impact of the implementation of the LUX formalism on the quark and gluon distributions. In CT18 LUX, the photon PDF determined directly by LUX QED is added to the nominal CT18 PDFs, with no adjustment of the momentum sum rule. In the standard CT18 QED proton, the momentum sum rule is enforced, but the momentum taken by the photon is subtracted from the sea quark distributions only. In the CT18 QED fit, the momentum distribution is determined by the fit itself.

We isolate the QED effects by comparing against QCD baselines with exactly the same setup (or as similar as possible) as the QCD+QED PDFs.
For NNPDF4.0, the QCD baseline was updated in the NNPDF4.0+QED paper to use a new theory pipeline~\cite{Barontini:2023vmr} which crucially meant recomputing all DIS with a new FONLL prescription~\cite{Barontini:2024xgu}.
In addition bugs in the implementation of a few datasets were found and corrected.
See appendix A of Ref.~\cite{NNPDF:2024djq} for more details and a comparison to the public NNPDF4.0 at NNLO. Isolating the QED effects in this way enables us to investigate and identify reasons for some of the differences observed in Table~\ref{tab:1}.

\begin{figure}[H]
\begin{center}
\includegraphics[scale=0.45,trim=0cm 0cm 0cm 1.395cm,clip]{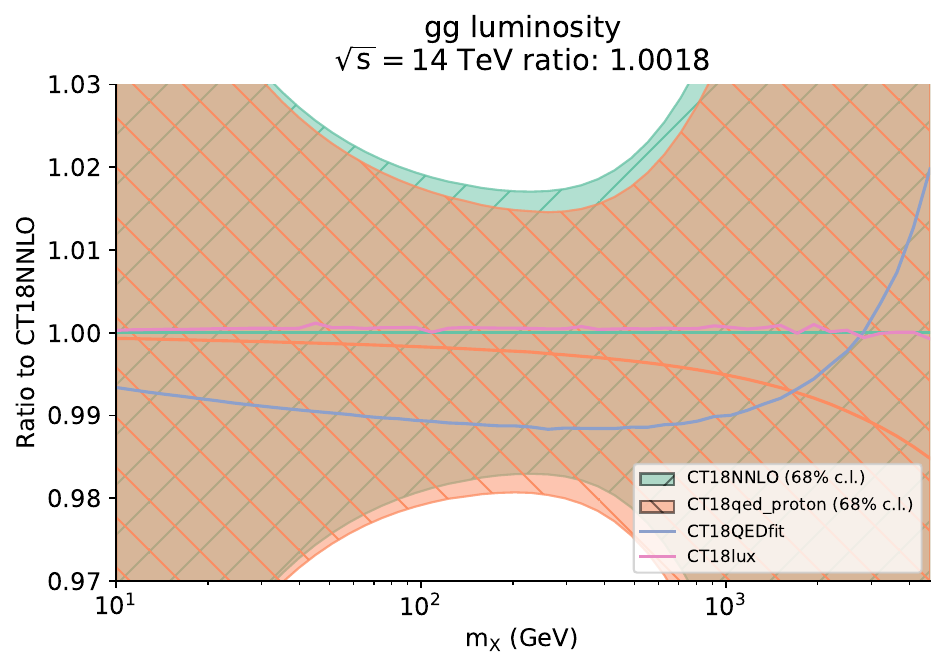}
\includegraphics[scale=0.45,trim=0cm 0cm 0cm 1.395cm,clip]{Figs/gglumi_MSHT20vsMSHT20QED_NNLO.pdf}
\includegraphics[scale=0.45,trim=0cm 0cm 0cm 1.395cm,clip]{Figs/gglumi_NNPDF3.1vsNNPDF3.1QED_NNLO.pdf}
\includegraphics[scale=0.45,trim=0cm 0cm 0cm 1.395cm,clip]{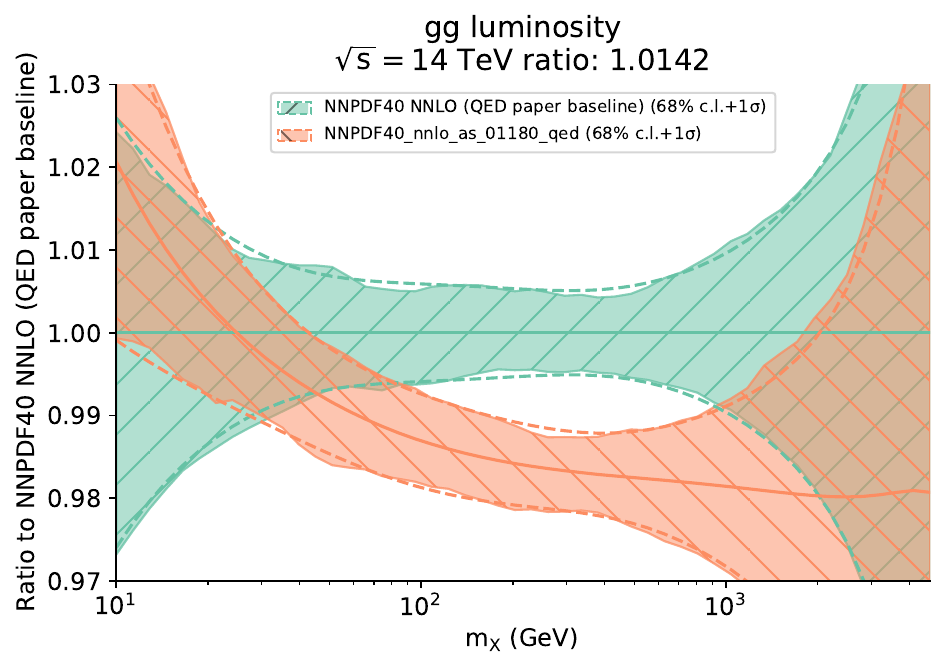}
\caption{\sf Comparison of the gluon-gluon luminosities at $\sqrt{s}=14~\rm{TeV}$ with NNLO PDFs. (Upper Left) CT18 in the QED proton, LUX and QED fit PDF sets relative to the QCD only CT18 baseline. (Upper Right) MSHT20 QED relative to QCD only. (Lower Left) NNPDF3.1 QED relative to a QCD only baseline. (Lower Right) NNPDF4.0 QED relative to a QCD only baseline. Note the MSHT and NNPDF3.1 plots are repeated from Figure~\ref{fig:gglumis_defaultNNLOQEDPDFs} as there are no changes to their QCD or QED comparison sets.}
\label{fig:gglumis_NNLOQEDPDFs_CT18luxqedfitNNPDF4.0updated}
\end{center}
\end{figure}

Firstly for CT18, the reason for the lack of change in the gluon momentum (and in turn the gluon-gluon luminosity) in the CT18+QED proton set is due to the first of the differences outlined above. That is due to how the photon momentum is reallocated from that of the QCD only fit partons. At NNLO, the LHAPDF CT18 LUX PDF and CT18+QED proton sets observe the smallest impacts on the gg luminosities. This occurs due to the lack of inclusion of the momentum sum rule impact in the former, and the by-hand subtraction of the required photon momentum from the quark sea in the latter, as noted in~\cite{Cooper-Sarkar:2025sqw}. This contrasts with the approach taken in MSHT, NNPDF  where the fit itself determines the parton momenta balance, i.e. from where the photon momentum is reallocated in the fit. Instead, for a direct comparison we take CT18+QED fit (as also used in~\cite{Cooper-Sarkar:2025sqw}), which follows the same strategy. In Figure~\ref{fig:gglumis_NNLOQEDPDFs_CT18luxqedfitNNPDF4.0updated}~(upper left) we add the CT18+QED fit and CT18 LUX PDFs to the gluon-gluon luminosity plot. It is clear that the behaviour is now more similar in shape as well as magnitude to that seen in NNPDF3.1 and MSHT.

Secondly, with the adoption of a more appropriate baseline\footnote{We note that some differences remain between the QCD+QED PDFs and QCD only PDFs here due to differences in the evolution, which we explore later.} for NNPDF4.0 at NNLO we find a reduction in the gg luminosity at $m_H$ of $-1.37\%$, in close agreement to that observed in NNPDF3.1, though still with a different shape (which we will explore further below), as shown in Figure~\ref{fig:gglumis_NNLOQEDPDFs_CT18luxqedfitNNPDF4.0updated}~(lower). Overall, we now find that MSHT20 see the smallest impacts on the gg luminosity ($-0.60\%$) at $m_H$, followed by CT18 (QED fit version) ($-1.09\%$) and NNPDF3.1 ($-1.36\%$ ), then NNPDF4.0 ($-1.37\%$) (shown in brackets in Table~\ref{tab:2} as it compares against the NNPDF4.0 QCD baseline set rather than the default LHAPDF NNPDF4.0). Focusing therefore on these PDFs, all groups observe that most of the momentum is taken from the gluon PDFs and the agreement becomes closer, as also seen in Figure~\ref{fig:gglumis_NNLOQEDPDFs_CT18luxqedfitNNPDF4.0updated}. 
Whilst these differences are small, and largely with the uncertainty bands of the PDFs, at the level of precision increasingly demanded of PDFs now and in the future, they are worthy of further study.

\begin{table}[h!]
\fontsize{8}{10}\selectfont 
  \renewcommand\arraystretch{1.1} 
\centering
\begin{tabular}{l C{1.0cm} C{1.2cm} C{2.05cm} C{2.25cm}}
\hline
\textbf{PDF set} & \textbf{Central} &
\textbf{\% error (symmetrised)} &
\textbf{\% change rel. to QCD only baseline at same order} &
\textbf{\% change rel. to NNLO QCD baseline} (3.1 for NNPDF) \\
\hline
MSHT20NNLO & 0.0206 & 1.16 & -- & -- \\
MSHT20NNLO+QED & 0.0205 & 1.09 & -0.60 & -0.60 \\
\hline
CT18 & 0.0205 & 1.93 & -- & -- \\
CT18+QED proton & 0.0205 & 1.93 & -0.19 & -0.19 \\
CT18 LUX & 0.0206 & 1.93 & 0.06 & 0.06 \\
CT18+QED fit~\textsuperscript{footnote }\tablefootnote{Note there is only the central PDF available for {\tt{CT18QEDfit}}.} & 0.0203 & -- & -1.09 & -1.09 \\
\hline
NNPDF3.1 & 0.0210 & 1.06 & -- & -- \\
NNPDF3.1+QED & 0.0207 & 0.74 & -1.36 & -1.36 \\
NNPDF4.0 & 0.0207 & 0.56 & -- & -1.44 \\
NNPDF4.0 QCD baseline~\textsuperscript{footnote }\tablefootnote{The corresponding PDF set used as the baseline is therefore taken from~\cite{NNPDF4NNLOQED_Website}.} & 0.0207 & 0.57 & -- & -1.70 \\
NNPDF4.0+QED & 0.0204 & 0.52 & -1.64 (-1.37) & -3.05 \\
\hline
\end{tabular}
\caption{PDF $gg$ luminosities for $\sqrt{s}=14$~TeV at $m_H=125$~GeV, focusing on the NNLO PDF sets. The bracketed result for NNPDF4.0+QED compares against their corresponding QCD baselines rather than the original public sets, for which there are slight differences. Removing these differences reduces slightly the QED impact observed.}
\label{tab:2}
\end{table}

\vspace{-0.5cm}

\begin{figure}[H]
\begin{center}
\includegraphics[scale=0.45,trim=0cm 0cm 0cm 1.395cm,clip]{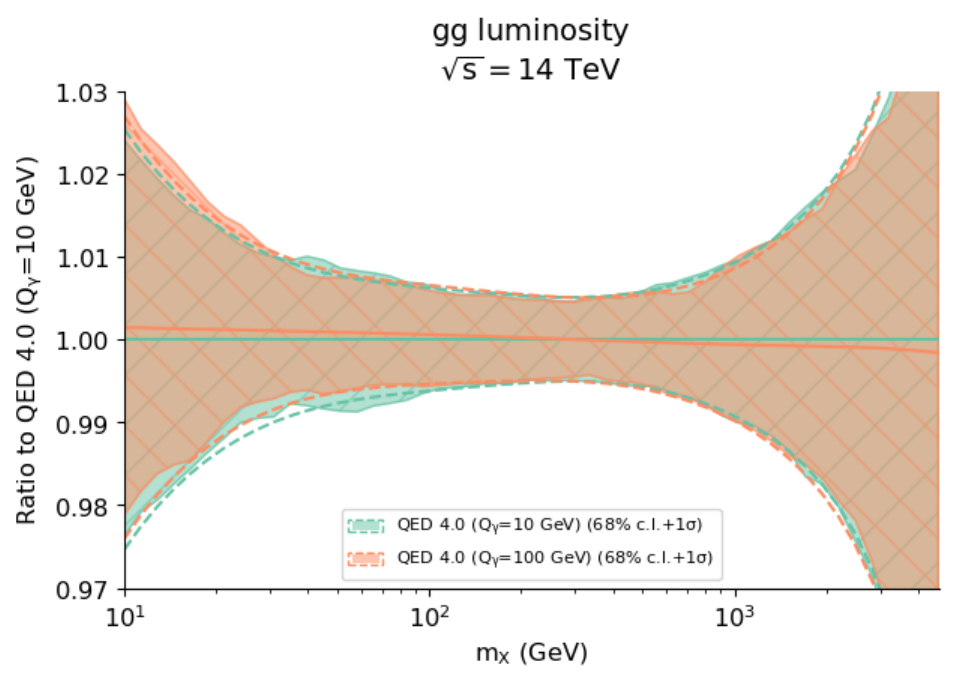}
\caption{\sf Comparison of the gluon-gluon luminosities at $\sqrt{s}=14~\rm{TeV}$ obtained when the photon is extracted at a starting scale of $Q_{\gamma} = 10 \rm{GeV}$ compared to the default $Q_{\gamma} = 100 \rm{GeV}$ in NNPDF4.0 QED.}
\label{fig:NNPDF4.0_diffphotonstartscale}
\end{center}
\end{figure}

The magnitude of the QED effect is largest in NNPDF, particularly in comparison to its smaller uncertainty band. 
One potential source of such differences has been suggested as the difference in the scale at which the photon is extracted via the LUXQED formalism, with MSHT and CT doing so at low scales ($Q_{\gamma} = 1, 1.3 {\rm GeV}$ respectively) and NNPDF doing so at large scales ($Q_{\gamma} = 100 {\rm GeV}$). To test this, in Figure~\ref{fig:NNPDF4.0_diffphotonstartscale} we compare in NNPDF4.0+QED the gluon-gluon luminosities obtained at $\sqrt{s}=14~{\rm TeV}$ with the photon PDF extracted instead at the lower scale of $Q_{\gamma} = 10 \rm{GeV}$, which is closer to that of CT and MSHT. As anticipated, this makes negligible difference as the full QCD$\otimes$QED evolution accounts for the change in scales, as also observed for the photon PDF in \cite{NNPDF:2024djq}.
For an exact comparison at $Q_{\gamma} \sim 1 \rm{GeV}$ it would be necessary to include uncertainties associated with the low $Q_{\gamma}$ which are not currently available in the NNPDF framework, but due to the lack of any effect seen when lowering the scale down to $Q_{\gamma} = 10 \rm{GeV}$ we consider further analysis at lower scales unnecessary at this stage.

Given the negligible impact of the starting scale, and the understood differences between CT18 PDF sets based upon different means of redistributing the QCD parton momenta to the photon, it appears that the remaining differences are at least partly down to aspects intrinsic to the different PDF sets. In our investigations  we therefore focus upon the differences between NNPDF3.1 and 4.0, as well as between the different 4.0 variants, where more direct comparisons can be made thanks to the availability of the public \texttt{nnpdf} framework~\cite{NNPDF:2021uiq}. We compare these, with the adoption of the appropriate QCD only baseline for 4.0, at NNLO in Figure~\ref{fig:gglumis_NNLOQEDPDFs_CT18luxqedfitNNPDF4.0updated}~(lower). In particular, whilst the magnitude of the impact of QED at $m_H$ is very similar between NNPDF3.1 in Figure~\ref{fig:gglumis_NNLOQEDPDFs_CT18luxqedfitNNPDF4.0updated}~(lower left) and 4.0 (with the adoption of the appropriate QCD only baseline for 4.0) in Figure~\ref{fig:gglumis_NNLOQEDPDFs_CT18luxqedfitNNPDF4.0updated}~(lower right) at NNLO, as noted in Table~\ref{tab:2}, it is clear the shape of the effect is still different. As a result, notable differences would still be observed in comparing NNPDF3.1 and 4.0 at NNLO at different $\sqrt{s}$ or invariant masses.

\begin{figure}[H]
\begin{center}
\includegraphics[scale=0.45,trim=0cm 0cm 0cm 1.395cm,clip]{Figs/figure30_plot_lumi1d.pdf}
\includegraphics[scale=0.45,trim=0cm 0cm 0cm 1.395cm,clip]{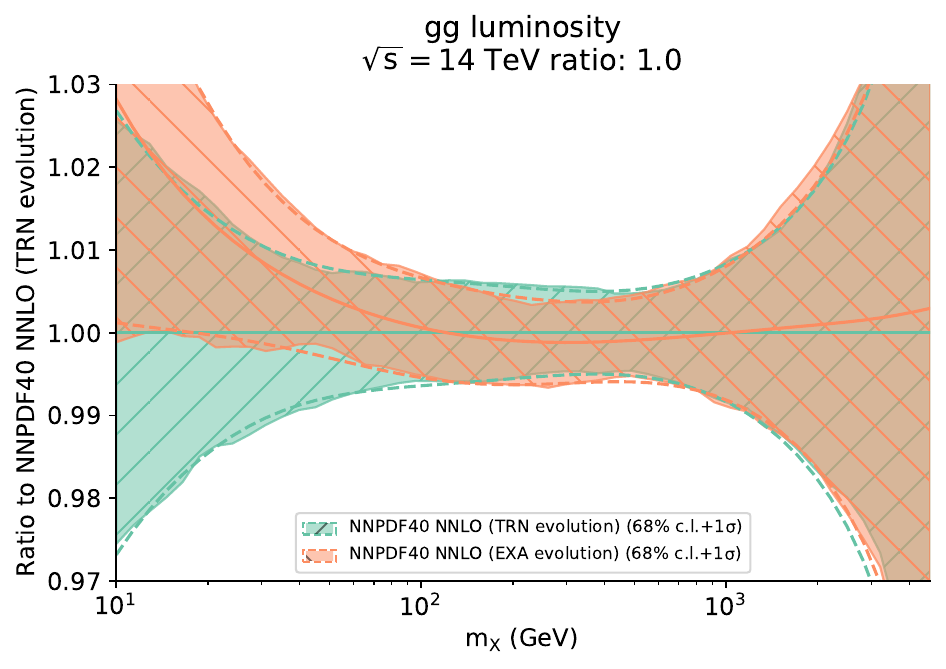}
\includegraphics[scale=0.45,trim=0cm 0cm 0cm 1.395cm,clip]{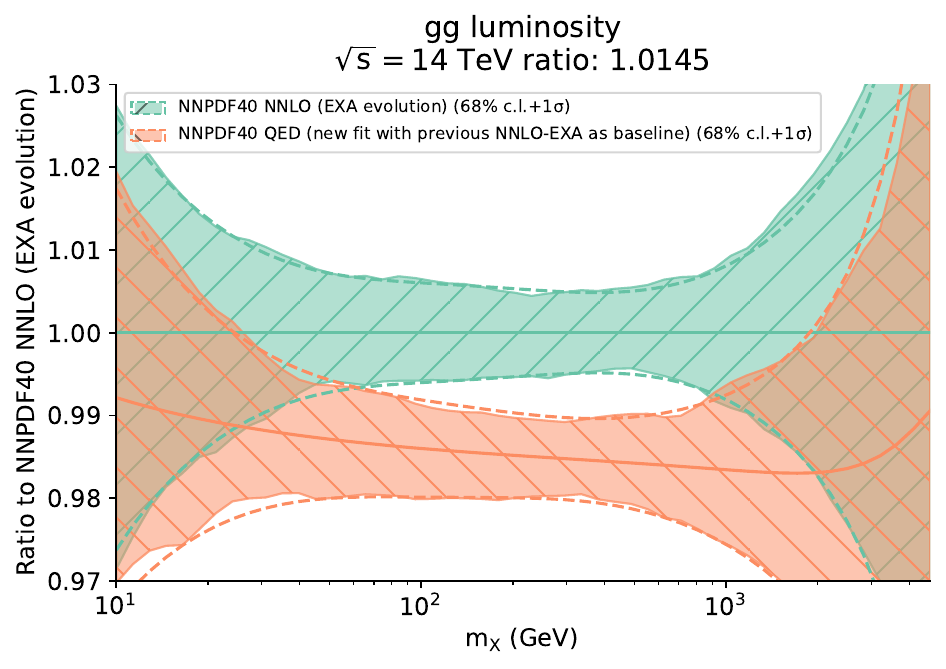}
\includegraphics[scale=0.45,trim=0cm 0cm 0cm 1.395cm,clip]{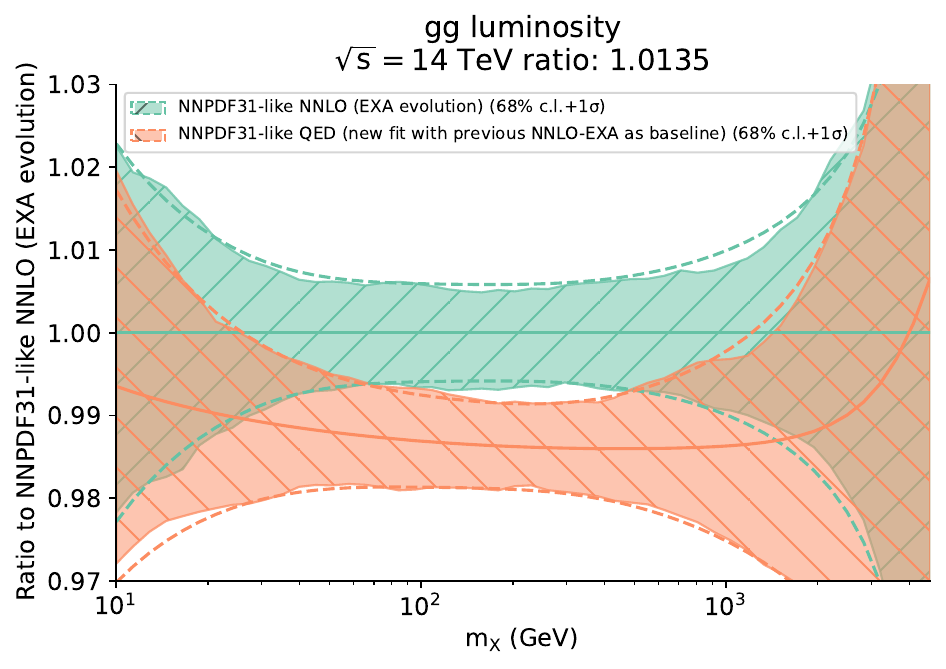}
\caption{\sf Comparison of the gluon-gluon luminosities at $\sqrt{s}=14~\rm{TeV}$ 
for a range of NNPDF NNLO fits. In the upper left figure, the QCD baseline corresponds to the baseline used in~\cite{NNPDF:2024djq}, in this paper the QCD fit is evolved using the truncated solution while the QED variant uses an exact solution as implemented in EKO~\cite{Candido:2022tld}. In the upper right figure, the same QCD only fit (4.0-like) is performed once with the exact solution (EXA) and once with truncated (TRN). In the lower left figure we therefore present a QCD/QED comparison in which both fits are evolved using the exact solution. While the effect of the QED corrections at the Higgs mass is almost unchanged, the difference in slope has almost completely disappeared. Finally, the lower right figure corresponds to the same pair of fits using instead a reduced dataset, closer to the dataset of NNPDF3.1.}
\label{fig:NNPDF4.0newQEDvsQCD}
\end{center}
\end{figure}

Further differences exist between the NNPDF3.1 and 4.0 fits, as well as between the baseline 4.0 QCD only and QCD+QED fits (shown in the last two rows in Table~\ref{tab:2} and in Figure~\ref{fig:gglumis_NNLOQEDPDFs_CT18luxqedfitNNPDF4.0updated}(lower)), that may explain some of this remaining difference.
With respect to NNPDF3.1, NNPDF4.0 has an enlarged dataset,
the fitting methodology was changed, integrability constraints were added to the fit, and the positivity is checked more strictly.
A detailed comparison can be found in~\cite{NNPDF:2021njg}.
Instead, with respect to the 4.0 QCD only fit, the QCD+QED fit included a change in the evolution settings, going from the truncated solution used also in 3.1~\cite{Bertone:2017bme} to the exact solution as implemented in Ref.~\cite{Candido:2022tld}.
This introduces a further difference when comparing the effect of the inclusion of QED in 3.1 (same settings in both fits) versus 4.0 (QCD and QCD+QED with different settings)\footnote{While the truncated solution offers advantages in N-space codes like EKO, it can only be implemented numerically when QED evolution is included with the subsequent loss of accuracy. See App. B of Ref~\cite{NNPDF:2024djq} for details.}.
We investigate a selection of these choices here, at NNLO (unless specified otherwise) we consider the no MHOU cases for 4.0 for closer comparison to 3.1.

In Figure~\ref{fig:NNPDF4.0newQEDvsQCD} we show explicitly the impact on the fit of some of these effects. We start in Figure~\ref{fig:NNPDF4.0newQEDvsQCD}~(upper left) with the current state of our comparison of the NNPDF NNLO QED fit with the QCD only baseline used in Ref~\cite{NNPDF:2024djq} (repeated from Figure~\ref{fig:gglumis_NNLOQEDPDFs_CT18luxqedfitNNPDF4.0updated}~(lower right)).
Next, we use the open-source NNPDF framework~\cite{NNPDF:2021uiq} to perform the same fit changing only the choice of evolution settings (truncated TRN vs exact EXA), both of them QCD only, shown in the upper right plot.
We see that the effect of the evolution is negligible at larger values of $m_{X}$ but becomes relevant for lower values, of order the uncertainty at lowest $m_{X}$ and changing the slope of the distribution.
The differences observed affect mainly small invariant masses as one might anticipate, since the difference between both approaches is formally of higher order and at lower invariant mass the perturbative expansion deteriorates.
In the lower left plot we therefore remove this difference from the comparison, we take our QCD+QED fit with exact evolution and now compared to its equivalent (in terms of both dataset and evolution) QCD only fit. The result is that the gap between both distributions remains (now being $-1.43\%$ at $m_H$) but the change in slope has now largely disappeared. The reduced slope with increasing $m_X$ is also more similar to the CT and MSHT results in Figure~\ref{fig:gglumis_NNLOQEDPDFs_CT18luxqedfitNNPDF4.0updated} (upper left and upper right), which is as expected as both use the exact evolution.

Finally, in the lower right figure the same fits are now repeated with a subset of data from the NNPDF4.0 fit so that it resembles the 3.1 dataset selection\footnote{Strictly speaking it is closer to 3.1.1' presented in PDF4LHC21~\cite{Cridge:2021qjj,PDF4LHCWorkingGroup:2022cjn}.}.
As was also found in~\cite{NNPDF:2021njg}, the main effect is of an increase on the uncertainties due to the smaller dataset but also a small reduction in the gap between the QCD only and QCD+QED fits at the relevant range for Higgs production (to $-1.25\%$ at $m_H$ and $\sqrt{s}=14~{\rm TeV}$).
In summary, this improves the comparison, making it much closer to the 3.1 case in Figure~\ref{fig:gglumis_defaultNNLOQEDPDFs}~(lower left) in shape whilst retaining a similar relative difference between the QCD+QED and QCD only central values. In turn it is also closer to the smaller QED impacts seen in CT and MSHT. Nonetheless, small differences in the precise magnitude of the effect remain.

In Table~\ref{tab:4} we provide a comparison of the gluon fusion Higgs total cross-sections calculated using the {\tt n3loxs} code~\cite{Baglio:2022wzu} for the same PDFs entering the gluon-gluon luminosity comparison in Table~\ref{tab:2}. As expected, the behaviour at the level of the luminosities follows through into the gluon fusion Higgs production cross-sections, as also noted in Table~2 of~\cite{Cooper-Sarkar:2025sqw} . 

\begin{table}[h!]
\fontsize{8}{10}\selectfont 
  \renewcommand\arraystretch{1.1} 
\centering
\begin{tabular}{l C{2.5cm} C{3.05cm}}
\hline
\textbf{PDF set} & \textbf{Central} &
\textbf{\% change rel. to QCD only baseline at same order}
\\
\hline
MSHT20NNLO & 51.78 & -- \\ 
MSHT20NNLO+QED & 51.48 & -0.57 \\ \hline
CT18 & 51.72 & -- \\ 
CT18+QED proton & 51.63 & -0.17\\ 
CT18 LUX & 51.75 & 0.06\\ 
CT18+QED fit & 51.17 & -1.08 \\ \hline
NNPDF3.1 & 52.90 & -- \\ 
NNPDF3.1+QED & 52.25 & -1.23 \\ 
NNPDF4.0 & 52.14 & -- \\ 
NNPDF4.0 QCD baseline & 52.01 & -- \\ 
NNPDF4.0+QED & 51.29 & -1.63 (-1.38) \\ 
\hline
\end{tabular}
\caption{Gluon Fusion Higgs production cross-sections [in $\mathrm{pb}$] at $\sqrt{s} = 14$ TeV, computed at NNLO with different NNLO PDF sets using \texttt{n3loxs}~\cite{Baglio:2022wzu}.}
\label{tab:4}
\end{table}

\subsection{Approximate N3LO}\label{sec:N3LO}

We can make similar comparisons at approximate N3LO, therefore in Table~\ref{tab:3} we expand the PDF sets shown in Table~\ref{tab:1} beyond the default ones to analyse the QED effects on the MSHT and NNPDF aN3LO PDF sets. In particular, we update the QCD only baseline PDFs to remove differences in settings relative to the QCD+QED PDFs.
In the MSHT case, this means taking the MSHT20aN3LO QCD baseline PDFs, which differ from the original MSHT20aN3LO PDFs by small dataset and other changes (see \cite{Cridge:2023ryv} for details).
Meanwhile, for NNPDF4.0 at aN3LO there are also slight differences between the public QCD and later QCD+QED PDFs. In order to ensure the underlying aN3LO ingredients are the same, we utilise the QCD baseline and QCD+QED results from the recent paper~\cite{Ball:2025xgq} for which the same settings are used for the QCD only and QCD+QED PDFs, these are the last two rows of Table~\ref{tab:3}. The comparisons against their QCD baselines rather than the public default QCD only PDFs are shown in brackets in Table~\ref{tab:3}.

\begin{figure}[H]
\begin{center}
\includegraphics[scale=0.45,trim=0cm 0cm 0cm 1.395cm,clip]{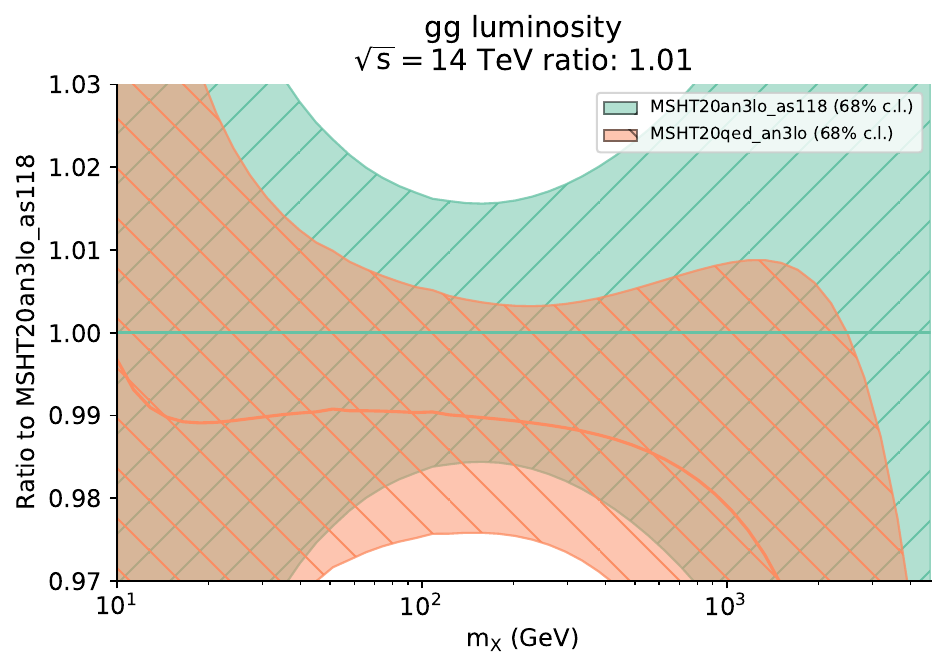}
\includegraphics[scale=0.45,trim=0cm 0cm 0cm 1.395cm,clip]{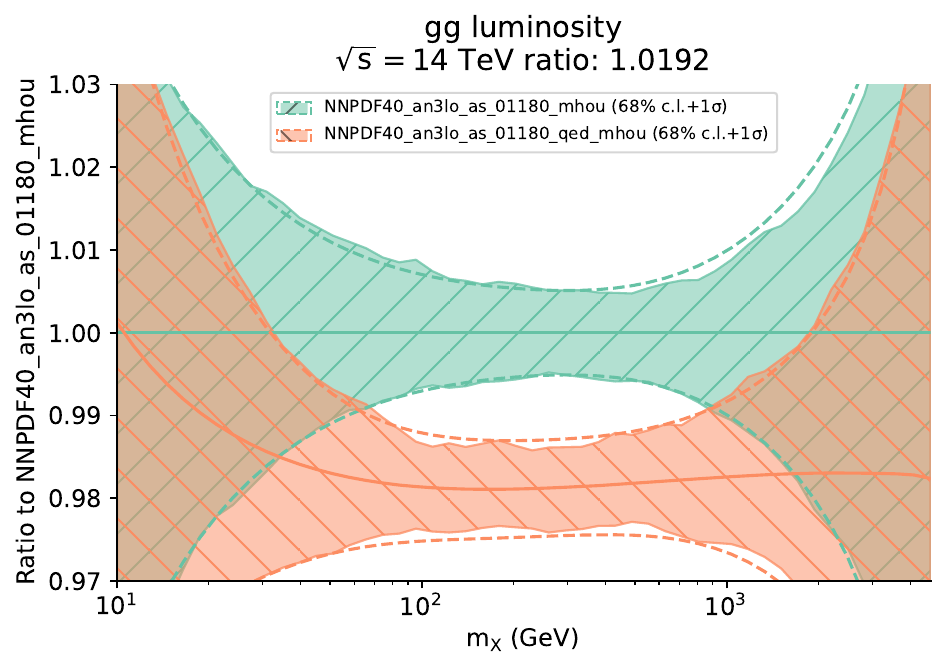}
\caption{\sf Comparison of the gluon-gluon luminosities at $\sqrt{s}=14~\rm{TeV}$ for the QCD+QED and QCD only PDFs of (left) MSHT20 and (right)  NNPDF4.0 at aN3LO using the PDFs available in LHAPDF.}
\label{fig:gglumis_defaultN3LOQEDPDFs}
\end{center}
\end{figure}

\begin{figure}[H]
\begin{center}
\includegraphics[scale=0.45,trim=0cm 0cm 0cm 1.395cm,clip]{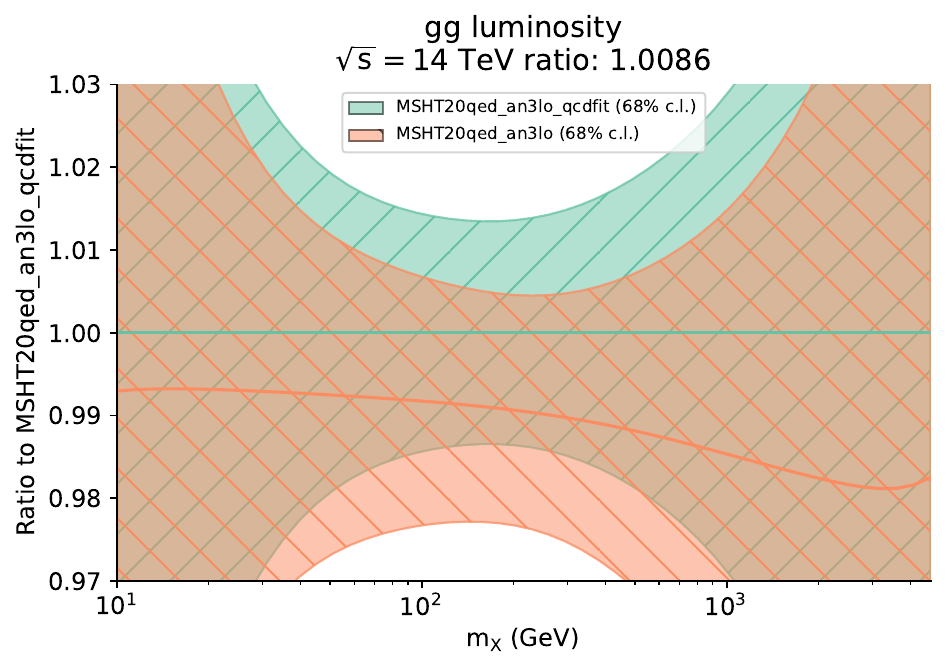}
\includegraphics[scale=0.45,trim=0cm 0cm 0cm 1.395cm,clip]{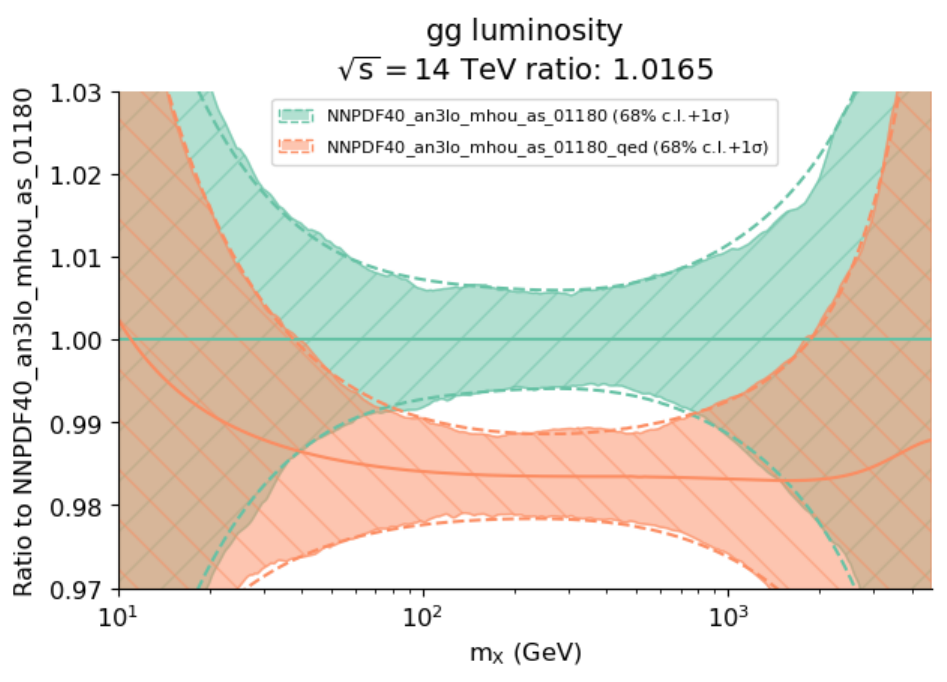}
\caption{\sf Comparison of the gluon-gluon luminosities at $\sqrt{s}=14~\rm{TeV}$ for the QCD+QED and QCD only PDFs of (left) MSHT20 and (right) NNPDF4.0 at aN3LO comparing QCD+QED PDFs against appropriate QCD only baselines.}
\label{fig:gglumis_N3LOQEDPDFs_correctQCDbaselines}
\end{center}
\end{figure}

As noted previously in~\cite{Cooper-Sarkar:2025sqw}, both MSHT and NNPDF4.0 observed slightly larger impacts of QED on the PDFs at aN3LO than at NNLO in their default PDF sets: $-0.99\%$ for MSHT, and $-1.88\%$ (or $-1.62\%$ in the without MHOU case not shown) for NNPDF. A comparison of their gluon-gluon luminosities over a range of invariant masses is provided in Figure~\ref{fig:gglumis_defaultN3LOQEDPDFs}. However, like at NNLO the QED impacts are slightly reduced by comparing with their QCD baselines (results shown in brackets in Table~\ref{tab:3}) to $-0.86\%$ for MSHT20 and $-1.57\%$ for NNPDF4.0 from~\cite{Ball:2025xgq} (compared to $-1.71\%$ against the public NNPDF4.0 aN3LO QCD only set). In the MSHT20 case the result is that the QED effect largely factorises from the QCD-order, with the QCD+QED/QCD PDF ratio being very similar at NNLO and aN3LO (though slightly larger in the latter), as noted in~\cite{Cridge:2023ryv}. In the case of NNPDF4.0 we utilise the aN3LO sets released together with the recent $\alpha_s$ determination of NNPDF~\cite{Ball:2025xgq}, for which the QCD+QED and QCD only fits truly correspond to the same baseline settings. We provide in Figure~\ref{fig:gglumis_N3LOQEDPDFs_correctQCDbaselines} this comparison of the QCD+QED PDFs against their identical QCD only baselines, observing slightly reduced differences relative to Figure~\ref{fig:gglumis_defaultN3LOQEDPDFs}. The reduction in the QED effect by comparing against the identical QCD only baseline is larger for NNPDF4.0 than for MSHT20. This therefore brings it closer to MSHT, though  differences still remain, with NNPDF4.0 still observing a larger impact of QED, like at NNLO.

\begin{figure}[H]
\begin{center}
\includegraphics[scale=0.45,trim=0cm 0cm 0cm 1.395cm,clip]{Figs/NNPDF4.0_an3lo_qedvsqcd.pdf}
\includegraphics[scale=0.45,trim=0cm 0cm 0cm 1.395cm,clip]{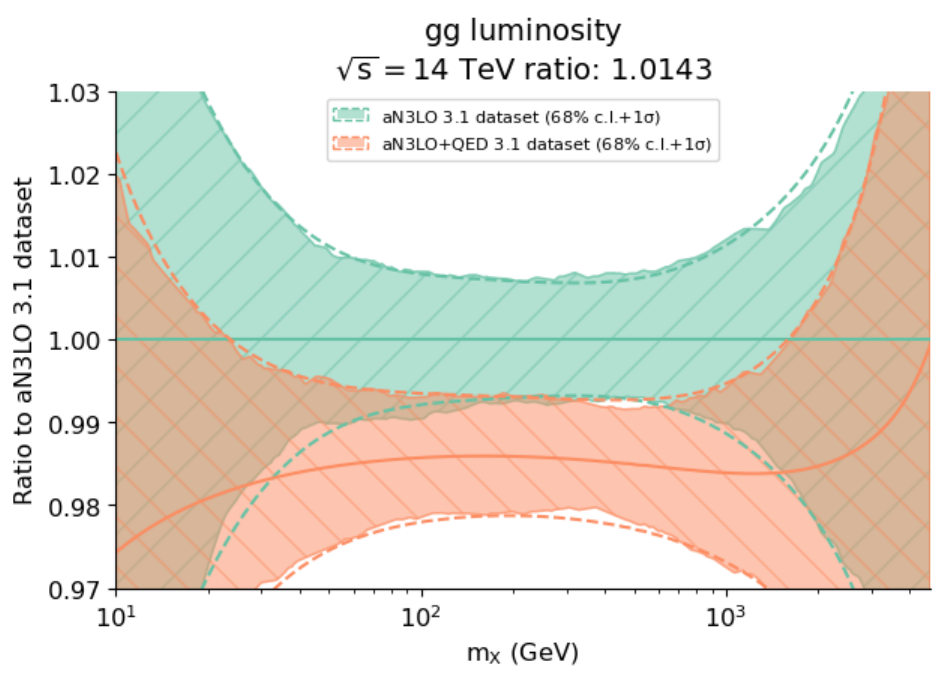}
\caption{\sf Comparison of the gluon-gluon luminosities at $\sqrt{s}=14~\rm{TeV}$ at aN3LO. (Left) NNPDF4.0 QCD+QED PDFs compared against the appropriate QCD only baseline (repeated from Figure~\ref{fig:gglumis_N3LOQEDPDFs_correctQCDbaselines}~(right)). (Right) Taking the 4.0 methodology with a 3.1-like dataset and comparing the QCD+QED PDFs against QCD only with other settings identical. Like in the NNLO case, the main difference between both fits regards an increase in the uncertainty of the fit and a small reduction of the gap.}
\label{fig:NNPDF4.0newQEDvsQCDN3LO}
\end{center}
\end{figure}

In Figure~\ref{fig:NNPDF4.0newQEDvsQCDN3LO}~(right) 
we show the change with the same reduced 3.1-like dataset utilised at NNLO, now at aN3LO, and again observe a reduced impact, with a luminosity change of $-1.41\%$ with the 3.1 like dataset (to be compared against $-1.57\%$ for the full 4.0 dataset). 
 As was the case at NNLO, we find that using the same baseline and a reduced dataset reduces the difference of the QCD+QED and QCD only PDFs. The simultaneous increase in the uncertainties from the reduced dataset also reduces the significance of the effect. This brings the results also closer to MSHT20 (see Figure~\ref{fig:gglumis_N3LOQEDPDFs_correctQCDbaselines}~(left)), but some difference in the magnitude of the effect remains.

\begin{table}[h!]
\fontsize{8}{10}\selectfont 
  \renewcommand\arraystretch{1.1} 
\centering
\begin{tabular}{l C{1.0cm} C{1.2cm} C{2.05cm} C{2.25cm}}
\hline
\textbf{PDF set} & \textbf{Central} &
\textbf{\% error (symmetrised)} &
\textbf{\% change rel. to QCD only baseline at same order} &
\textbf{\% change rel. to NNLO QCD baseline} (4.0 for NNPDF) \\
\hline
MSHT20aN3LO & 0.0195 & 1.71 & -- & -5.20 \\
MSHT20aN3LO QCD baseline~\textsuperscript{footnote }\tablefootnote{Note this is the PDF set {\tt{MSHT20qed\_aN3LO\_qcdfit}}, which is the QCD only baseline for the aN3LO QCD + QED fit, produced for~\cite{Cridge:2023ryv} and available on the MSHT website~\cite{MSHT_website}.}  & 0.0195 & 1.42 & -- & -5.33 \\
MSHT20aN3LO+QED & 0.0193 & 1.47 & -0.99 (-0.86) & -6.14 \\
\hline
NNPDF4.0aN3LO (inc. MHOU)~\textsuperscript{footnote }\tablefootnote{We take the NNPDF4.0aN3LO PDF sets including missing higher order uncertainties (MHOU) as these are the default and most comparable to MSHT20aN3LO where such MHOU are also included.} & 0.0202 & 0.65 & -- & -2.45 \\
NNPDF4.0aN3LO+QED (inc. MHOU) & 0.0198 & 0.55 & -1.88 & -4.25 \\
NNPDF4.0aN3LO (inc. MHOU) QCD baseline~\textsuperscript{footnote }\tablefootnote{We take this QCD baseline set from~\cite{Ball:2025xgq} to ensure the same settings are used amongst the QCD only and QCD+QED sets for this comparison. It is called {\tt NNPDF40\_an3lo\_mhou\_as\_01180} and its QCD+QED equivalent has {\tt \_qed} after in the name. It is available at~\cite{NNPDF4aN3LOQEDalphas_Website}} & 0.0202 & 0.67 & -- & -2.59 \\
NNPDF4.0aN3LO+QED (inc. MHOU) cf QCD baseline & 0.0199 & 0.52 & -1.71 (-1.57) & -4.12 \\
\hline
\end{tabular}
\caption{PDF $gg$ luminosities for $\sqrt{s}=14$~TeV at $m_H=125$~GeV for the aN3LO PDFs. Note the most relevant QCD only baseline at the same order is taken, e.g. including or not including MHOU for NNPDF depending on the set being compared. The bracketed results for MSHT20aN3LO+QED and NNPDF4.0+QED compare against their corresponding QCD baselines rather than the original public sets, for which there are slight differences. Removing these differences reduces slightly the QED impact observed.}
\label{tab:3}
\end{table}

For completeness in Table~\ref{tab:5} we again show the gluon fusion Higgs total cross-sections calculated using the {\tt n3loxs} code~\cite{Baglio:2022wzu} for the same PDFs for which we provided the gluon-gluon luminosities in Table~\ref{tab:3}. Once more, as expected the same behaviour is seen in the gluon fusion Higgs production cross-sections as detailed in the comparison of the gluon-gluon luminosities\footnote{We remind the reader that there is an increase in the Higgs cross-section from the N3LO hard cross-section relative to that at NNLO, which is somewhat compensated by the reduction in the gluon PDFs, as noted in~\cite{McGowan:2022nag,NNPDF:2024nan,Cridge:2024icl}.} and as also seen in Table~2 of~\cite{Cooper-Sarkar:2025sqw}.

\begin{table}[h!]
\fontsize{8}{10}\selectfont 
  \renewcommand\arraystretch{1.1} 
\centering
\begin{tabular}{l C{2.5cm} C{3.05cm}}
\hline
\textbf{PDF set} & \textbf{Central} &
\textbf{\% change rel. to QCD only baseline at same order}
\\
\hline
MSHT20aN3LO & 50.86 & -- \\
MSHT20aN3LO QCD baseline & 50.78 & -- \\
MSHT20aN3LO+QED & 50.35 & -1.00 (-0.85) \\ \hline
NNPDF40aN3LO (inc. MHOU) & 52.64 & -- \\
NNPDF40aN3LO+QED (inc. MHOU) & 51.64 & -1.90 \\
NNPDF40aN3LO (inc. MHOU) QCD baseline & 52.58 & -- \\
NNPDF40aN3LO+QED (inc. MHOU) cf QCD baseline & 51.72 & -1.75 (-1.62) \\
\hline
\end{tabular}
\caption{Gluon Fusion Higgs production cross-sections [in $\mathrm{pb}$] at $\sqrt{s} = 14$ TeV, computed at N3LO with different aN3LO PDF sets using \texttt{n3loxs}~\cite{Baglio:2022wzu}.}
\label{tab:5}
\end{table}

Finally, we comment that it is noteworthy also that across PDF groups, over time the gluon-gluon luminosity at the Higgs mass has been reducing, due to a combination of the effects of the inclusion of new data (e.g. NNPDF3.1 to NNPDF4.0), QED effects (seen across all groups) and approximate N3LO (MSHT and NNPDF)\footnote{We also note that the recent MSHT~\cite{Harland-Lang:2025wvm} and NNPDF~\cite{Ball:2025xtj} analyses of higher twist corrections also observe slight reductions in the gluon PDF and/or gluon-gluon luminosities around $m_H$. Separately CT25~\cite{Ablat:2025gdb,Ablat:2025gbp} also observe a slight reduction in the gluon in the Higgs region relative to CT18A.}. The combination of aN3LO+QED in MSHT20 results in a net reduction of $-6.14\%$ (see Table~\ref{tab:1} and Table~\ref{tab:3}), in comparison the change in NNPDF4.0 aN3LO+QED (with MHOU) is $-4.0\%$ compared to the 4.0 QCD baseline ($-4.25\%$ compared to the public NNPDF4.0) (see Table~\ref{tab:3}). As there was a reduction in gluon-gluon luminosity at $m_H$ from NNPDF3.1 to NNPDF4.0, NNPDF4.0 aN3LO+QED with MHOU is reduced by $-5.63\%$ at $m_H$ compared to the NNPDF3.1 result (see Table~\ref{tab:1}). Updates to the approximate N3LO PDFs (not discussed here), in terms of both the determination of further Mellin moments of the splitting functions~\cite{Falcioni:2023luc,Falcioni:2023vqq,Falcioni:2023tzp,Falcioni:2024xyt,Falcioni:2024qpd} and calculation of transition matrix elements~\cite{Ablinger:2022wbb,Ablinger:2023ahe,Ablinger:2024xtt}, further bring the results at aN3LO closer together. MSHT observe an increase of $\lesssim 2\%$ in the gluon-gluon luminosity around $m_H$~\cite{Cridge:2025oel} and NNPDF observe a slight reduction of $\approx 0.5\%$\cite{Cridge:2024icl} relative to their respective public aN3LO baselines.

\section{Conclusions}

Overall, this initial work highlights that several of the differences in the impacts of QED across PDF global fitting groups can be understood. In particular those due to different choices in the implementation of the QED effects and subtleties therein. These include how the QCD parton momenta is reallocated to the photon PDF in the QCD+QED case, which explains the differences relative to CT18 QED+proton.
In addition, subtle differences in settings and baselines between the QCD only and QCD+QED PDFs can spuriously modify the observed differences.
This is in particular the case for NNPDF4.0, where using a different baseline and evolution settings, alters the shape of the PDF upon inclusion of QED effects, artificially enlarging the difference.

Even after all differences have been removed, and QED effects are included consistently, sub-percent level discrepancies between the different PDFs remain when comparing the QCD+QED fits to their QCD only counterparts.
These remaining differences appear to arise somewhat out of inherent differences in the PDFs themselves (also present in the QCD only PDFs) which may be exacerbated by the inclusion of the photon.
As an example, we find that reducing the NNPDF4.0 dataset to be closer to NNPDF3.1 (on top of removing the differences in evolution and QCD baseline in the former) reduces the impact of QED on the gluon-gluon luminosity at $m_H$, bringing the results somewhat closer to those of CT and MSHT. Whilst remaining differences between groups are typically within uncertainties, they may be relevant in the future and further studies will be required to identify their sources. 

A benchmarking of the inclusion of QED effects across the groups may be the only way to explore remaining differences between groups in full detail. This would have the supplementary benefit of allowing the inclusion of the additional uncertainty from the slight differences in the QED effects observed across different PDFs to also be accounted for in the uncertainty, for example on the Higgs production cross-section. Currently only the important effect on the central value is incorporated via the recommendation in \cite{Cooper-Sarkar:2025sqw} as a full combination would be necessary to include the effect of these differences.

\section*{Acknowledgments}
 We wish to thank the authors of the paper \cite{Cooper-Sarkar:2025sqw} for the impetus to perform these further investigations, as well as for their useful comments. We also are grateful to Lucian Harland-Lang, Robert Thorne and Felix Hekhorn for useful discussions and comments. We thank the organisers of the Les Houches  2025 (``Physics at TeV colliders and Beyond the Standard Model'') Standard Model Session, where this work began. We also thank PDF4LHC members for useful discussions and members of the LHC Higgs Working Group for discussions that initiated the larger study. T.C. acknowledges that this work has been supported by funding from Research Foundation-Flanders (FWO) (application number: 12E1323N) and by the Royal Society through Grant URF$\backslash$R1$\backslash$251540. J.C-M. acknowledges funding from the Ramón y Cajal program grant RYC2023-043794-I funded by MCIN/AEI/10.13039/501100011033 and by ESF+.

\bibliography{references}{}

@article{Bailey:2020ooq,
    author = "Bailey, S. and Cridge, T. and Harland-Lang, L. A. and Martin, A. D. and Thorne, R. S.",
    title = "{Parton distributions from LHC, HERA, Tevatron and fixed target data: MSHT20 PDFs}",
    eprint = "2012.04684",
    archivePrefix = "arXiv",
    primaryClass = "hep-ph",
    reportNumber = "IPPP/20/58",
    doi = "10.1140/epjc/s10052-021-09057-0",
    journal = "Eur. Phys. J. C",
    volume = "81",
    number = "4",
    pages = "341",
    year = "2021"
}

@article{NNPDF:2021njg,
    author = "Ball, Richard D. and others",
    collaboration = "NNPDF",
    title = "{The path to proton structure at 1\% accuracy}",
    eprint = "2109.02653",
    archivePrefix = "arXiv",
    primaryClass = "hep-ph",
    reportNumber = "Edinburgh 2021/12, Nikhef-2021-013, TIF-UNIMI-2021-11",
    doi = "10.1140/epjc/s10052-022-10328-7",
    journal = "Eur. Phys. J. C",
    volume = "82",
    number = "5",
    pages = "428",
    year = "2022"
}

@article{Barontini:2023vmr,
    author = "Barontini, Andrea and Candido, Alessandro and Cruz-Martinez, Juan M. and Hekhorn, Felix and Schwan, Christopher",
    title = "{Pineline: Industrialization of high-energy theory predictions}",
    eprint = "2302.12124",
    archivePrefix = "arXiv",
    primaryClass = "hep-ph",
    reportNumber = "TIF-UNIMI-2023-4, CERN-TH-2023-021",
    doi = "10.1016/j.cpc.2023.109061",
    journal = "Comput. Phys. Commun.",
    volume = "297",
    pages = "109061",
    year = "2024"
}

@article{Barontini:2024xgu,
    author = "Barontini, Andrea and Candido, Alessandro and Hekhorn, Felix and Magni, Giacomo and Stegeman, Roy",
    title = "{An FONLL prescription with coexisting flavor number PDFs}",
    eprint = "2408.07383",
    archivePrefix = "arXiv",
    primaryClass = "hep-ph",
    reportNumber = "Nikhef-2024-014, Edinburgh 2024/5, TIF-UNIMI-2024-13",
    doi = "10.1007/JHEP10(2024)004",
    journal = "JHEP",
    volume = "10",
    pages = "004",
    year = "2024"
}

@article{Candido:2022tld,
    author = "Candido, Alessandro and Hekhorn, Felix and Magni, Giacomo",
    title = "{EKO: evolution kernel operators}",
    eprint = "2202.02338",
    archivePrefix = "arXiv",
    primaryClass = "hep-ph",
    reportNumber = "TIF-UNIMI-2022-2, Nikhef-2022-003",
    doi = "10.1140/epjc/s10052-022-10878-w",
    journal = "Eur. Phys. J. C",
    volume = "82",
    number = "10",
    pages = "976",
    year = "2022"
}

@article{NNPDF:2021uiq,
    author = "Ball, Richard D. and others",
    collaboration = "NNPDF",
    title = "{An open-source machine learning framework for global analyses of parton distributions}",
    eprint = "2109.02671",
    archivePrefix = "arXiv",
    primaryClass = "hep-ph",
    reportNumber = "Edinburgh 2021/13, Nikhef-2021-020, TIF-UNIMI-2021-12",
    doi = "10.1140/epjc/s10052-021-09747-9",
    journal = "Eur. Phys. J. C",
    volume = "81",
    number = "10",
    pages = "958",
    year = "2021"
}

@article{Hou:2019efy,
    author = "Hou, Tie-Jiun and others",
    title = "{New CTEQ global analysis of quantum chromodynamics with high-precision data from the LHC}",
    eprint = "1912.10053",
    archivePrefix = "arXiv",
    primaryClass = "hep-ph",
    reportNumber = "MSUHEP-19-025, PITT-PACC-1911, SMU-HEP-19-03",
    doi = "10.1103/PhysRevD.103.014013",
    journal = "Phys. Rev. D",
    volume = "103",
    number = "1",
    pages = "014013",
    year = "2021"
}

@article{ATLAS:2021vod,
    author = "Aad, Georges and others",
    collaboration = "ATLAS",
    title = "{Determination of the parton distribution functions of the proton using diverse ATLAS data from $pp$ collisions at $\sqrt{s} = 7$, 8 and 13~TeV}",
    eprint = "2112.11266",
    archivePrefix = "arXiv",
    primaryClass = "hep-ex",
    reportNumber = "CERN-EP-2021-239",
    doi = "10.1140/epjc/s10052-022-10217-z",
    journal = "Eur. Phys. J. C",
    volume = "82",
    number = "5",
    pages = "438",
    year = "2022"
}

@article{Alekhin:2017kpj,
    author = {Alekhin, S. and Bl\"umlein, J. and Moch, S. and Placakyte, R.},
    title = "{Parton distribution functions, $\alpha_s$, and heavy-quark masses for LHC Run II}",
    eprint = "1701.05838",
    archivePrefix = "arXiv",
    primaryClass = "hep-ph",
    reportNumber = "DESY-16-179, DO-TH-16-13",
    doi = "10.1103/PhysRevD.96.014011",
    journal = "Phys. Rev. D",
    volume = "96",
    number = "1",
    pages = "014011",
    year = "2017"
}

@article{McGowan:2022nag,
    author = "McGowan, J. and Cridge, T. and Harland-Lang, L. A. and Thorne, R. S.",
    title = "{Approximate N$^{3}$LO parton distribution functions with theoretical uncertainties: MSHT20aN$^3$LO PDFs}",
    eprint = "2207.04739",
    archivePrefix = "arXiv",
    primaryClass = "hep-ph",
    doi = "10.1140/epjc/s10052-023-11236-0",
    journal = "Eur. Phys. J. C",
    volume = "83",
    number = "3",
    pages = "185",
    year = "2023"
}

@article{NNPDF:2024nan,
    author = "Ball, Richard D. and others",
    collaboration = "NNPDF",
    title = "{The Path to N$^3$LO Parton Distributions}",
    eprint = "2402.18635",
    archivePrefix = "arXiv",
    primaryClass = "hep-ph",
    reportNumber = "Nikhef-2023-020, TIF-UNIMI-2023-23, Edinburgh 2023/29",
    month = "2",
    year = "2024"
}

@article{NNPDF:2024dpb,
    author = "Ball, Richard D. and others",
    collaboration = "NNPDF",
    title = "{Determination of the theory uncertainties from missing higher orders on NNLO parton distributions with percent accuracy}",
    eprint = "2401.10319",
    archivePrefix = "arXiv",
    primaryClass = "hep-ph",
    reportNumber = "TIF-UNIMI-2023-22, Edinburgh 2023/34, CERN-TH-2024-009",
    doi = "10.1140/epjc/s10052-024-12772-z",
    journal = "Eur. Phys. J. C",
    volume = "84",
    number = "5",
    pages = "517",
    year = "2024"
}

@article{NNPDF:2017mvq,
    author = "Ball, Richard D. and others",
    collaboration = "NNPDF",
    title = "{Parton distributions from high-precision collider data}",
    eprint = "1706.00428",
    archivePrefix = "arXiv",
    primaryClass = "hep-ph",
    reportNumber = "EDINBURGH-2017-08, NIKHEF-2017-006, OUTP-17-04P, TIF-UNIMI-2017-3, CAVENDISH-HEP-17-06, CERN-TH-2017-077, Edinburgh 2017/08,
  Nikhef/2017-006, OUTP-17-04P,TIF-UNIMI-2017-3",
    doi = "10.1140/epjc/s10052-017-5199-5",
    journal = "Eur. Phys. J. C",
    volume = "77",
    number = "10",
    pages = "663",
    year = "2017"
}

@article{Bertone:2017bme,
    author = "Bertone, Valerio and Carrazza, Stefano and Hartland, Nathan P. and Rojo, Juan",
    collaboration = "NNPDF",
    title = "{Illuminating the photon content of the proton within a global PDF analysis}",
    eprint = "1712.07053",
    archivePrefix = "arXiv",
    primaryClass = "hep-ph",
    reportNumber = "Nikhef/2017-064, CERN-TH-2017-235, NIKHEF-2017-064",
    doi = "10.21468/SciPostPhys.5.1.008",
    journal = "SciPost Phys.",
    volume = "5",
    number = "1",
    pages = "008",
    year = "2018"
}

@article{Alekhin:2012ig,
    author = "Alekhin, S. and Blumlein, J. and Moch, S.",
    title = "{Parton Distribution Functions and Benchmark Cross Sections at NNLO}",
    eprint = "1202.2281",
    archivePrefix = "arXiv",
    primaryClass = "hep-ph",
    reportNumber = "DESY-12-023, DO-TH-11-31, LPN-12-033, SFB-CPP-12-08",
    doi = "10.1103/PhysRevD.86.054009",
    journal = "Phys. Rev. D",
    volume = "86",
    pages = "054009",
    year = "2012"
}

@article{Cerutti:2025yji,
    author = "Cerutti, Matteo and Accardi, Alberto and Fernando, Ishara P. and Li, Shujie and Owens, Joseph F. and Park, Sanghwa",
    title = "{Systematic uncertainties from higher-twist corrections in DIS at large x}",
    eprint = "2501.06849",
    archivePrefix = "arXiv",
    primaryClass = "hep-ph",
    reportNumber = "JLAB-THY-25-7",
    doi = "10.1103/PhysRevD.111.094013",
    journal = "Phys. Rev. D",
    volume = "111",
    number = "9",
    pages = "094013",
    year = "2025"
}

@article{Harland-Lang:2025wvm,
    author = "Harland-Lang, L. A. and Cridge, T. and Reader, M. and Thorne, R. S.",
    title = "{A reassessment of the role of high $x$ data on the MSHT global PDF fit}",
    eprint = "2510.03753",
    archivePrefix = "arXiv",
    primaryClass = "hep-ph",
    month = "10",
    year = "2025"
}

@article{Ball:2025xtj,
    author = "Ball, Richard D. and Chiefa, Amedeo and Stegeman, Roy",
    title = "{Parton distributions with higher twist and jet power corrections}",
    eprint = "2511.14387",
    archivePrefix = "arXiv",
    primaryClass = "hep-ph",
    reportNumber = "Edinburgh 2025/29",
    month = "11",
    year = "2025"
}

@article{Ball:2025xgq,
    author = "Ball, Richard D. and Barontini, Andrea and Cruz-Martinez, Juan and Forte, Stefano and Hekhorn, Felix and Nocera, Emanuele R. and Rojo, Juan and Stegeman, Roy",
    collaboration = "NNPDF",
    title = "{A determination of $\alpha _s(m_Z)$ at ${{\textrm{aN}}}^3{{\textrm{LO}}}_{{\textrm{QCD}}}\otimes {{\textrm{NLO}}}_{{\textrm{QED}}}$ accuracy from a global PDF analysis}",
    eprint = "2506.13871",
    archivePrefix = "arXiv",
    primaryClass = "hep-ph",
    reportNumber = "TIF-UNIMI-2025-10, Edinburgh 2024/12, CERN-TH-2025-106",
    doi = "10.1140/epjc/s10052-025-14676-y",
    journal = "Eur. Phys. J. C",
    volume = "85",
    number = "9",
    pages = "1001",
    year = "2025"
}

@article{Xie:2021equ,
    author = "Xie, Keping and Hobbs, T. J. and Hou, Tie-Jiun and Schmidt, Carl and Yan, Mengshi and Yuan, C. -P.",
    collaboration = "CTEQ-TEA",
    title = "{Photon PDF within the CT18 global analysis}",
    eprint = "2106.10299",
    archivePrefix = "arXiv",
    primaryClass = "hep-ph",
    reportNumber = "MSUHEP-21-013, PITT-PACC-2112, FERMILAB-PUB-21-370-QIS-SCD-T, SMU-HEP-21-06, SMU-HEP-21-06,
  FERMILAB-PUB-21-370-QIS-SCD-T",
    doi = "10.1103/PhysRevD.105.054006",
    journal = "Phys. Rev. D",
    volume = "105",
    number = "5",
    pages = "054006",
    year = "2022"
}

@article{NNPDF:2024djq,
    author = "Ball, Richard D. and others",
    collaboration = "NNPDF",
    title = "{Photons in the proton: implications for the LHC}",
    eprint = "2401.08749",
    archivePrefix = "arXiv",
    primaryClass = "hep-ph",
    reportNumber = "TIF-UNIMI-2023-17, Edinburgh 2023/19, CERN-TH-2023-159",
    doi = "10.1140/epjc/s10052-024-12731-8",
    journal = "Eur. Phys. J. C",
    volume = "84",
    number = "5",
    pages = "540",
    year = "2024"
}

@misc{NNPDF4NNLOQED_Website,
    key = "https://nnpdf.mi.infn.it/wp-content/uploads/2024/05/NNPDF40_nnlo_as_01180_qcd.tar.gz",
    note = "NNPDF4.0 Website: \url{https://nnpdf.mi.infn.it/wp-content/uploads/2024/05/NNPDF40\_nnlo\_as\_01180\_qcd.tar.gz}" 
}

@misc{MSHT_Website,
    key = "https://www.hep.ucl.ac.uk/msht/Grids/MSHT20qed_an3lo_qcdfit.tar.gz",
    note = "MSHT20 Website: \url{https://www.hep.ucl.ac.uk/msht/Grids/MSHT20qed_an3lo_qcdfit.tar.gz}" 
}

@article{Cridge:2021pxm,
    author = "Cridge, T. and Harland-Lang, L. A. and Martin, A. D. and Thorne, R. S.",
    title = "{QED parton distribution functions in the MSHT20 fit}",
    eprint = "2111.05357",
    archivePrefix = "arXiv",
    primaryClass = "hep-ph",
    reportNumber = "IPPP/21/44",
    doi = "10.1140/epjc/s10052-022-10028-2",
    journal = "Eur. Phys. J. C",
    volume = "82",
    number = "1",
    pages = "90",
    year = "2022"
}

@article{Cridge:2023ryv,
    author = "Cridge, Thomas and Harland-Lang, Lucian A. and Thorne, Robert S.",
    title = "{Combining QED and approximate ${\rm N}^3$LO QCD corrections in a global PDF fit: MSHT20qed\_an3lo PDFs}",
    eprint = "2312.07665",
    archivePrefix = "arXiv",
    primaryClass = "hep-ph",
    reportNumber = "DESY-23-195",
    doi = "10.21468/SciPostPhys.17.1.026",
    journal = "SciPost Phys.",
    volume = "17",
    number = "1",
    pages = "026",
    year = "2024"
}

@inproceedings{Barontini:2024dyb,
    author = "Barontini, Andrea and Laurenti, Niccolo and Rojo, Juan",
    title = "{NNPDF4.0 aN$^3$LO PDFs with QED corrections}",
    booktitle = "{31st International Workshop on Deep-Inelastic Scattering and Related Subjects}",
    eprint = "2406.01779",
    archivePrefix = "arXiv",
    primaryClass = "hep-ph",
    month = "6",
    year = "2024"
}

@article{Xie:2023qbn,
    author = "Xie, Keping and Zhou, Bei and Hobbs, T. J.",
    collaboration = "CTEQ-TEA",
    title = "{The photon content of the neutron}",
    eprint = "2305.10497",
    archivePrefix = "arXiv",
    primaryClass = "hep-ph",
    reportNumber = "ANL-182626, MSUHEP-24-004, PITT-PACC-2314",
    doi = "10.1007/JHEP04(2024)022",
    journal = "JHEP",
    volume = "04",
    pages = "022",
    year = "2024"
}

@article{Manohar:2016nzj,
    author = "Manohar, Aneesh and Nason, Paolo and Salam, Gavin P. and Zanderighi, Giulia",
    title = "{How bright is the proton? A precise determination of the photon parton distribution function}",
    eprint = "1607.04266",
    archivePrefix = "arXiv",
    primaryClass = "hep-ph",
    reportNumber = "CERN-TH-2016-155",
    doi = "10.1103/PhysRevLett.117.242002",
    journal = "Phys. Rev. Lett.",
    volume = "117",
    number = "24",
    pages = "242002",
    year = "2016"
}

@article{Manohar:2017eqh,
    author = "Manohar, Aneesh V. and Nason, Paolo and Salam, Gavin P. and Zanderighi, Giulia",
    title = "{The Photon Content of the Proton}",
    eprint = "1708.01256",
    archivePrefix = "arXiv",
    primaryClass = "hep-ph",
    reportNumber = "CERN-TH-2017-141",
    doi = "10.1007/JHEP12(2017)046",
    journal = "JHEP",
    volume = "12",
    pages = "046",
    year = "2017"
}

@article{Cridge:2021qjj,
    author = "Cridge, Thomas",
    collaboration = "PDF4LHC21 combination group",
    title = "{PDF4LHC21: Update on the benchmarking of the CT, MSHT and NNPDF global PDF fits}",
    eprint = "2108.09099",
    archivePrefix = "arXiv",
    primaryClass = "hep-ph",
    doi = "10.21468/SciPostPhysProc.8.101",
    journal = "SciPost Phys. Proc.",
    volume = "8",
    pages = "101",
    year = "2022"
}

@article{PDF4LHCWorkingGroup:2022cjn,
    author = "Ball, Richard D. and others",
    collaboration = "PDF4LHC Working Group",
    title = "{The PDF4LHC21 combination of global PDF fits for the LHC Run III}",
    eprint = "2203.05506",
    archivePrefix = "arXiv",
    primaryClass = "hep-ph",
    reportNumber = "Edinburgh 2021/31, FERMILAB-PUB-22-121-QIS-SCD-T, MSUHEP-22-010, SMU-HEP-22-01, Nikhef 2021-033",
    doi = "10.1088/1361-6471/ac7216",
    journal = "J. Phys. G",
    volume = "49",
    number = "8",
    pages = "080501",
    year = "2022"
}

@article{Cooper-Sarkar:2025sqw,
    author = "Cooper-Sarkar, Amanda M. and Cridge, Thomas and Hobbs, T. J. and Huston, Joey and Nadolsky, Pavel and Ponce-Chavez, Maximiliano and Xie, Keping",
    title = "{QED-enhanced PDF implications for the Higgs sector}",
    eprint = "2508.06603",
    archivePrefix = "arXiv",
    primaryClass = "hep-ph",
    reportNumber = "ANL-198026, MSUHEP-25-017",
    month = "8",
    year = "2025"
}

@article{Cridge:2024icl,
    author = "Cridge, Thomas and others",
    title = "{Combination of aN$^3$LO PDFs and implications for Higgs production cross-sections at the LHC}",
    eprint = "2411.05373",
    archivePrefix = "arXiv",
    primaryClass = "hep-ph",
    reportNumber = "DESY-24-134, TIF-UNIMI-2024-17, Edinburgh 2024/9, CERN-TH-2024-167",
    doi = "10.1088/1361-6471/adde78",
    journal = "J. Phys. G",
    volume = "52",
    pages = "6",
    year = "2025"
}

@inproceedings{Cridge:2025oel,
    author = "Cridge, T. and Harland-Lang, L. A. and Thorne, R. S.",
    title = "{MSHT Approximate N3LO PDFs: Updates and Consequences for Phenomenology}",
    booktitle = "{32nd International Workshop on Deep-Inelastic Scattering and Related Subjects}",
    eprint = "2510.09321",
    archivePrefix = "arXiv",
    primaryClass = "hep-ph",
    month = "10",
    year = "2025"
}

@article{Falcioni:2023luc,
    author = "Falcioni, G. and Herzog, F. and Moch, S. and Vogt, A.",
    title = "{Four-loop splitting functions in QCD \textendash{} The quark-quark case}",
    eprint = "2302.07593",
    archivePrefix = "arXiv",
    primaryClass = "hep-ph",
    reportNumber = "DESY 23--022, LTH 1333",
    doi = "10.1016/j.physletb.2023.137944",
    journal = "Phys. Lett. B",
    volume = "842",
    pages = "137944",
    year = "2023"
}

@article{Falcioni:2023vqq,
    author = "Falcioni, G. and Herzog, F. and Moch, S. and Vogt, A.",
    title = "{Four-loop splitting functions in QCD \textendash{} The gluon-to-quark case}",
    eprint = "2307.04158",
    archivePrefix = "arXiv",
    primaryClass = "hep-ph",
    reportNumber = "DESY 23-096, LTH 1345",
    doi = "10.1016/j.physletb.2023.138215",
    journal = "Phys. Lett. B",
    volume = "846",
    pages = "138215",
    year = "2023"
}

@article{Falcioni:2023tzp,
    author = "Falcioni, G. and Herzog, F. and Moch, S. and Vermaseren, J. and Vogt, A.",
    title = "{The double fermionic contribution to the four-loop quark-to-gluon splitting function}",
    eprint = "2310.01245",
    archivePrefix = "arXiv",
    primaryClass = "hep-ph",
    reportNumber = "ZU-TH 62/23, DESY 23-146, Nikhef 2023-015, LTH 1353",
    doi = "10.1016/j.physletb.2023.138351",
    journal = "Phys. Lett. B",
    volume = "848",
    pages = "138351",
    year = "2024"
}

@article{Falcioni:2024xyt,
    author = "Falcioni, G. and Herzog, F. and Moch, S. and Pelloni, A. and Vogt, A.",
    title = "{Four-loop splitting functions in QCD \textendash{} The quark-to-gluon case}",
    eprint = "2404.09701",
    archivePrefix = "arXiv",
    primaryClass = "hep-ph",
    reportNumber = "ZU-TH 20/24, DESY-24-053, LTH 1367",
    doi = "10.1016/j.physletb.2024.138906",
    journal = "Phys. Lett. B",
    volume = "856",
    pages = "138906",
    year = "2024"
}

@article{Falcioni:2024qpd,
    author = "Falcioni, G. and Herzog, F. and Moch, S. and Pelloni, A. and Vogt, A.",
    title = "{Four-loop splitting functions in QCD \textendash{} the gluon-gluon case \textendash{}}",
    eprint = "2410.08089",
    archivePrefix = "arXiv",
    primaryClass = "hep-ph",
    reportNumber = "ZU-TH 47/24, DESY-24-144, LTH 1384",
    doi = "10.1016/j.physletb.2024.139194",
    journal = "Phys. Lett. B",
    volume = "860",
    pages = "139194",
    year = "2025"
}

@article{Ablinger:2022wbb,
    author = {Ablinger, J. and Behring, A. and Bl\"umlein, J. and De Freitas, A. and Goedicke, A. and von Manteuffel, A. and Schneider, C. and Sch\"onwald, K.},
    title = "{The unpolarized and polarized single-mass three-loop heavy flavor operator matrix elements A$_{gg,Q}$ and \ensuremath{\Delta}A$_{gg,Q}$}",
    eprint = "2211.05462",
    archivePrefix = "arXiv",
    primaryClass = "hep-ph",
    reportNumber = "DESY 15-112, DO-TH 22/26, CERN-TH-2022-179, ZU-TH 53/22, RISC Report
  Series 22-25, MSUHEP-22-036",
    doi = "10.1007/JHEP12(2022)134",
    journal = "JHEP",
    volume = "12",
    pages = "134",
    year = "2022"
}

@article{Ablinger:2023ahe,
    author = {Ablinger, J. and Behring, A. and Bl\"umlein, J. and De Freitas, A. and von Manteuffel, A. and Schneider, C. and Sch\"onwald, K.},
    title = "{The first\textendash{}order factorizable contributions to the three\textendash{}loop massive operator matrix elements AQg(3) and \ensuremath{\Delta}AQg(3)}",
    eprint = "2311.00644",
    archivePrefix = "arXiv",
    primaryClass = "hep-ph",
    reportNumber = "DO-TH 23/12, DESY 23-142, CERN-TH-2023-164, MSUHEP-23-025, RISC Report series 23-12, ZU-TH 60/23",
    doi = "10.1016/j.nuclphysb.2023.116427",
    journal = "Nucl. Phys. B",
    volume = "999",
    pages = "116427",
    year = "2024"
}

@article{Ablinger:2024xtt,
    author = {Ablinger, J. and Behring, A. and Bl\"umlein, J. and De Freitas, A. and von Manteuffel, A. and Schneider, C. and Sch\"onwald, K.},
    title = "{The non-first-order-factorizable contributions to the three-loop single-mass operator matrix elements $A_{Qg}^{(3)}$ and $\Delta A_{Qg}^{(3)}$}",
    eprint = "2403.00513",
    archivePrefix = "arXiv",
    primaryClass = "hep-ph",
    reportNumber = "DO--TH 23/15. DESY 24--027, RISC Report series 24--02, ZU-TH 13/24,
  CERN-TH-2024-30, DO-TH 23/15, RISC Report series 24--02, ZU-TH 13/24,
  CERN-TH-2024-30, DESY-24-027",
    month = "3",
    year = "2024"
}

@inproceedings{Ablat:2025gdb,
    author = "Ablat, A. and others",
    title = "{CT25: Progress toward next-generation PDFs for precision phenomenology at the LHC}",
    eprint = "2512.19779",
    archivePrefix = "arXiv",
    primaryClass = "hep-ph",
    reportNumber = "ANL-201496, MSUHEP-25-026",
    month = "12",
    year = "2025"
}

@article{Ablat:2025gbp,
    author = "Ablat, Alim and Dulat, Sayipjamal and Guzzi, Marco and Huston, Joey and Mohan, Kirtimaan and Nadolsky, Pavel and Stump, Dan and Yuan, C. -P.",
    title = "{Strong Coupling Constant Determination from the new CTEQ-TEA Global QCD Analysis}",
    eprint = "2512.23792",
    archivePrefix = "arXiv",
    primaryClass = "hep-ph",
    reportNumber = "MSUHEP-25-019",
    month = "12",
    year = "2025"
}

@article{Baglio:2022wzu,
    author = "Baglio, Julien and Duhr, Claude and Mistlberger, Bernhard and Szafron, Robert",
    title = "{Inclusive production cross sections at N$^{3}$LO}",
    eprint = "2209.06138",
    archivePrefix = "arXiv",
    primaryClass = "hep-ph",
    reportNumber = "CERN-TH-2022-109, SLAC-PUB-17699, BONN-TH-2022-22",
    doi = "10.1007/JHEP12(2022)066",
    journal = "JHEP",
    volume = "12",
    pages = "066",
    year = "2022"
}

@misc{NNPDF4aN3LOQEDalphas_Website,
    key = "https://data.nnpdf.science/pdfs/NNPDF40_an3lo_mhou_as_01180.tar.gz",
    note = "NNPDF4.0 Website: \url{hhttps://data.nnpdf.science/pdfs/NNPDF40_an3lo_mhou_as_01180.tar.gz}" 
}
\bibliographystyle{h-physrev}

\end{document}